\documentclass[preprint2]{emulateapj}

\shorttitle{Nulling beam combiners}
\shortauthors{Guyon et al.}
\usepackage{graphicx}
\usepackage[mathscr]{eucal}
\begin{document}

\title{Optimal beam combiner design for nulling interferometers}

\author{Olivier Guyon}
\affil{Steward Observatory, University of Arizona, Tucson, AZ 85721, USA}
\affil{National Astronomical Observatory of Japan, Subaru Telescope, Hilo, HI 96720}
\email{guyon@naoj.org}

\author{Bertrand Mennesson}
\author{Eugene Serabyn}
\author{Stefan Martin}
\affil{Jet Propulsion Laboratory, 4800 Oak Grove Drive, Pasadena, CA 91109, USA}

\begin{abstract}

A scheme to optimally design a beam combiner is discussed for any pre-determined fixed geometry nulling interferometer aimed at detection and characterization of exoplanets with multiple telescopes or a single telescope (aperture masking). We show that considerably higher order nulls can be achieved with 1-D interferometer geometries than possible with 2-D geometries with the same number of apertures. Any 1-D interferometer with N apertures can achieve a $2(N-1)$-order null, while the order of the deepest null for a random 2-D aperture geometry interferometer is the order of the N-th term in the Taylor expansion of $e^{i(x^2+y^2)}$ around x=0, y=0 ($2^{nd}$ order null for $N=2,3$; $4^{th}$ order null for $N=4,5,6$). We also show that an optimal beam combiner for nulling interferometry relies only 0 or $\pi$ phase shifts. Examples of nulling interferometer designs are shown to illustrate these findings.

\end{abstract}

\section{Introduction}
\label{sec:intro}

Spectral characterization of Earth-like planets around other stars may reveal the presence of life, and is therefore of high scientific value. Acquiring high quality spectra of small rocky planets in the habitable zones around nearby stars requires an instrument that can optically separate starlight from planet light in order to avoid being limited by photon noise from the host star. Two approaches have been studied in the last few decades: nulling interferometry (usually with an array of telescopes), and single aperture coronagraphy. A nulling interferometer is an interferometer designed to cancel light from an on-axis source (usually a star) while keeping as much as possible of the light from faint sources close to the central star. Nulling interferometers can thus detect thermal emission from exoplanets \citep{1978Natur.274..780B}, and are particularly attractive at infrared wavelengths (about 5 to 30 $\mu$m) for which the planet to star contrast is more favorable than in visible light. At this wavelength, a single aperture (+coronagraph) option would require a large telescope due to the linear dependence of angular resolution with wavelength, and an interferometer consisting of widely separated telescopes is a more suitable approach \citep{1979Icar...38..136B,1998ARA&A..36..507W,2007STIN...0814326L,2009AsBio...9....1C}. Nulling interferometers with short (few meter) baselines have also been proposed for visible light observations of exoplanets \citep{2010ASPC..430..368S}.

A key limitation of the two-telescope nulling interferometer proposed by \cite{1978Natur.274..780B} is due to the finite angular size of stars, which, even in the absence of instrumental defects, makes it impossible to fully cancel starlight while preserving light from a faint nearby source (planet). With a 2-telescope configuration, the ideal nulling interferometer throughput for a point source is proportional to the square of the angular separation to the optical axis: the interferometer produces a second order null near the optical axis, commonly referred to as a $\theta^2$ null, where $\theta$ is the angular separation to the optical axis. Since stellar diameters are typically about 1 percent of the planet to star separation for a system similar to the Earth-Sun system, the maximum differential attenuation between starlight and planet light attainable with a 2-telescope design is therefore around $10^4$, short of the $\approx 10^6$ (thermal IR) to $\approx 10^{10}$ (visible light) contrast between the two objects. 

To overcome this limitation, nulling interferometers with more than two apertures have been proposed to achieve higher order nulls, thus offering better extinction on partially resolved stellar disks. The extinction is then a function of both the interferometer geometry and the interferometric combination between the aperture beams \citep{1999ASPC..194..423L}, and many interferometer designs have been proposed, with increasing null depth (quantified by the null order). The Angel Cross design \citep{1990ngst.conf...81A} combines for example two Bracewell interferometers in a 2-D geometry to achieve a $\theta^4$ null. \cite{1997ApJ...475..373A} later showed that a linear 4-aperture design can offer a $\theta^6$ null. With 5 telescopes, a solution offering a $\theta^8$ null was also proposed \citep{1997ASPC..119..285W}. \cite{1996Icar..123..249L,1997Icar..128..202M} established array geometry requirements to reach a given nulling order, and proposed a 5-aperture solution offering a deep null and able to distinguish the signal from a planet from a symetrical exozodiacal cloud. \cite{2003ESASP.539..565R} showed that arbitrary null orders can be obtained, given a sufficient number of apertures, and discusses in a later paper \citep{2006dies.conf..213R} the practical usefulness of higher null orders, confirming that interferometers with higher order nulls are more resilient to phase errors \citep{2003ApOpt..42.1867M}.

Null depth is unfortunately often achieved with more complex interferometers at the expense of efficiency: for a fixed number of apertures, a smaller fraction of the total light gathered by the interferometer is used as null order is increased. For example, in the 4-aperture Angel Cross design, 25\% of the light is used (only one of the four interferometer outputs offers the $\theta^4$ null), while a simpler 2-aperture Bracewell offers 50\% throughput with a $\theta^2$ null. Null depth and throughput must therefore be balanced to find the optimal interferometer design when total mission cost/complexity are taken into account, and the scientific return of the mission must be maximized. Consequently, the optimal array geometry may sacrifice a very deep null in favor of other factors such as maintaining high throughput, required phase shifts in the beam combiner \citep{2002ApOpt..41.4697M} or more easily realized aperture geometries, such as arranging all all apertures on a circle to avoid long delay lines. For example, the Darwin mission study adopted a 4-aperture geometry on a circle, with a $2^{nd}$ order null \citep{2009ExA....23..435C}, with a 50\% efficiency (two of the four interferometer outputs are used for science), and for which stellar leakage is about 100 times brighter than light from an Earth-like planet at 10pc. Simple 3-aperture configurations have also been explored \citep{2004SPIE.5491.1639S, 2004SPIE.5491..831K}.

It is therefore important to understand how the achievable null depth and throughput are constrained by aperture geometry, to simultaneously optimize the aperture geometry and beam combiner design. This interferometer optimization problem however remains largely unsolved, as the development of new nulling interferometer designs has consisted of series of point designs, with no single design method leading to improvements, and no clear understanding of where the performance limits might be, given array geometry constraints. The relative importance of array geometry and beam combiner design is especially unclear, as previously published nulling interferometer designs rely on a particular match between aperture positions and beam combinations to achieve high null orders, making it impossible to decouple their relative impact on performance. The goal of this paper is to provide a universal method to optimally design the beam combiner for any nulling interferometer geometry, and thus to establish performance limits of nulling interferometry given realistic array constraints (such as maximum number of telescope, or maximum baseline). To achieve this goal, a universal mathematical model of the interferometer is first established and described in Section \ref{sec:model}. This model is then used in Section \ref{sec:optc} to derive, for a given entrance aperture geometry, the optimal beam combiner design. Nulling interferometer design examples are given in Section \ref{sec:designs} to illustrate the findings of this paper. We remind the reader that the analysis presented in this paper is purely based on a photon-noise limited detection assumption, and does not take into account manufacturing constraints, calibration issues, and image synthesis performance (which is an important criterion when the planet is embedded in an exozodiacal cloud).

\section{Nulling interferometer model}
\label{sec:model}

\subsection{Relationship with Coronagraphy}

Traditionally, coronagraphs use a single aperture pupil (which may be composed of adjacent segments) and perform starlight rejection with masks introducing amplitude and/or phase modifications in pupil and/or focal planes. Nulling interferometers use a sparse array of telescopes, and perform starlight rejection by coherent destructive interference between the beams. The boundaries between the two techniques have become less clear as schemes using nulling interferometry schemes on a single aperture exist (see for example \cite{2000A&AS..141..319B,2010ASPC..430..480K,2011ApJ...743..178M}) or using coronagraphic techniques on sparse apertures \citep{2001A&A...370..680A,2002A&A...391..379G,2002A&A...396..345R}.

For this paper, it is assumed that an interferometer is defined by its sparse entrance aperture (which may be obtained by aperture masking of a single larger aperture), while a coronagraph is a nulling device on a single aperture. No distinction is made between the types of nulling devices: coronagraphs using masks in pupil and focal planes, or nulling interferometers using coherent combinations of a finite set of beams. In an earlier publication \citep{2006ApJS..167...81G}, the fundamental limits of coronagraph performance for high contrast imaging were derived using a model of the coronagraph akin to a nulling interferometer. The telescope entrance pupil was split into a set of subapertures, which were coherently combined to produce output beams simultaneously achieving high starlight rejection and good transmission for planet light. This approach is justified by linearity in complex amplitude for both coronagraphs and interferometers, which leads to equivalence between the two approaches: for a finite field of view, a coronagraph can be modeled as a set of coherent interferences between a finite set of subpupils paving the telescope entrance aperture. Thanks to this equivalence, a universal algebraic representation of nulling devices (including coronagraphs) could be used to derive the fundamental limits of coronagraphy, using linearity in complex amplitude as the only constraint to the performance.

The approach used in \cite{2006ApJS..167...81G} is therefore equally applicable to both coronagraphs (nulling device on a single aperture) and interferometers (nulling device on sparse aperture). When this approach is used on sparse apertures, as done in this paper, it can directly give, for a given aperture geometry, the optimal design for the nulling device: which beams should be coherently mixed together, with the corresponding mixing ratios and phase shifts. In Section \ref{ssec:algebra}, this algebraic modeling approach is described and adapted to sparse apertures.

\subsection{Nulling interferometer algebraic representation}
\label{ssec:algebra}

\subsubsection{Aperture Geometry}

In this study, it is assumed that the source observed by the interferometer is unresolved by individual apertures: there is full coherence within an aperture. The nulling interferometer design is then fully described by its aperture geometry (number, sizes and positions of the telescopes in the interferometer) and the coherent interferometric combinations performed between the apertures. With N the number of apertures, each aperture k is described by its 2-D position $(x_k,y_k)$ on the plane normal to the line of sight and its radius $r_k$, which can be defined as the square root of the collecting area of the aperture divided by $\sqrt{\pi}$ if the aperture is not circular. The aperture geometry is thus fully described by N, and $(x_k,y_k,r_k)$, with $k=0..N-1$.

\subsubsection{Nulling interferometer model, and intensity response for an unresolved point source}

\begin{figure*}
\includegraphics[scale=0.4]{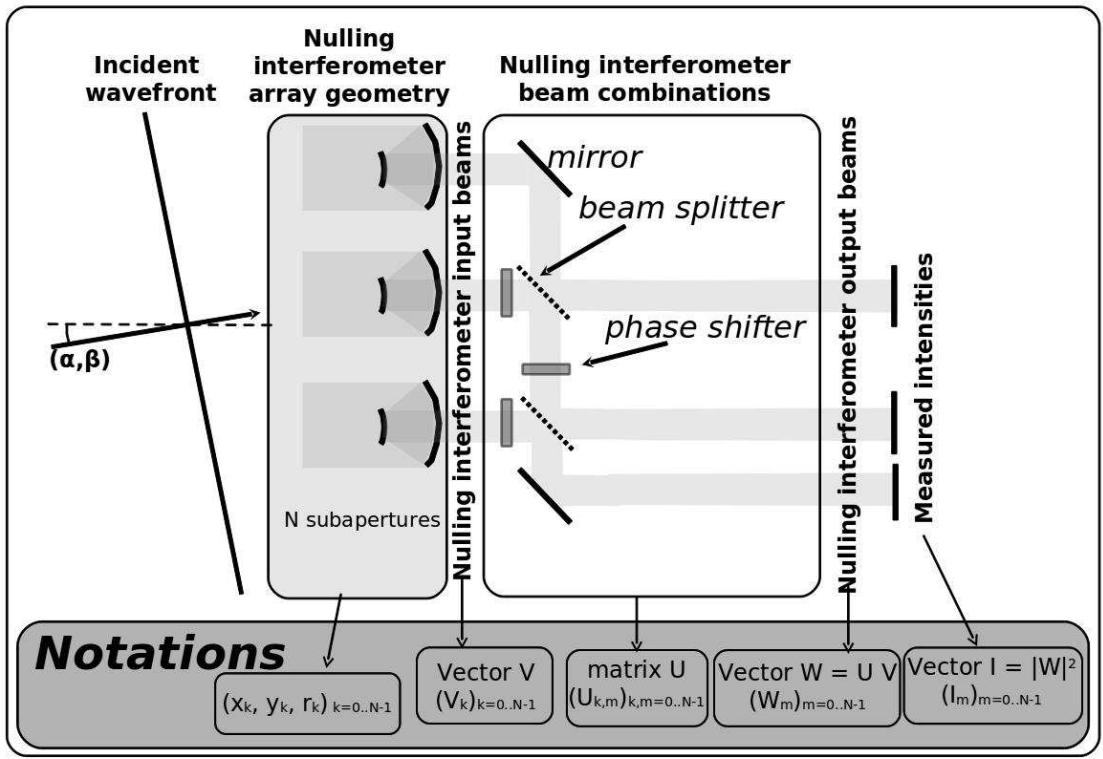} 
\caption{\label{fig:notations} Mathematical representation of a nulling interferometer and corresponding Notations. The indices used for the inputs and outputs of the interferometer's beam combination unit are $n$ and $m$ respectively. The nulling beam combiner, which physically consists of beam splitters and phase shifters, is represented as a complex values matrix $U$ which links the beam combiner input (vector $V$ of complex values) to its output (vector $W$ of complex values). Note that the arrangement of beam splitters, phase shifter and mirrors shown in this figure may not be representative of a real beam combiner, which may adopt a different arrangement of beam splitters. Any beam combiner can however be represented as a matrix $U$.}
\end{figure*}

The mathematical representation of a nulling interferometer used in this paper is illustrated in Figure \ref{fig:notations}. The position of a point source at infinity is defined by its angular offset $(\alpha,\beta)$ from the interferometer optical axis. For this point source, the N-element complex amplitude vector V which describes the set of complex amplitudes at the entrance of the interferometer's apertures is:

\begin{equation}
\label{equ:vectvk}
V_k = r_k e^{i 2 \pi (x_k \alpha + y_k \beta)/\lambda} 
\end{equation}

The interferometric combinations performed between the apertures are fully described by a NxN complex matrix $U$ which links the interferometer's outputs to its inputs (= vector V). Linearity in complex amplitude imposes that the outputs are represented by a complex amplitude vector W which is a linear function of the input vector V:
\begin{equation}
W = U V
\end{equation}

The science detector measures the square modulus $I = |W|^2$ of W. The matrix $U$ represents the design of the nulling device, and is not a function of the input complex amplitudes V. Coefficients of $U$ are complex numbers denoted $U_{k,m}$, with k the aperture index and m the interferometer output index. 
Each column $U_k$ of $U$ records how light from a single input $k$ (a single subaperture) is directed to all outputs. Each row $U_m$ of $U$ (m fixed) records the complex conjugates of the set of input complex amplitudes which, when "fed" into the interferometer, will send all of the light into interferometer output m. Since this set of inputs can generally not be written as a vector V according to equation \ref{equ:vectvk}, there may not be a position on the sky for which all light would be directed to output m (although one might choose the matrix $U$ to send all light from a given sky position $(\alpha,\beta)$ to a single output m by setting $U_m$ to be the complex conjugate of $V(\alpha,\beta)$). Since $|V|^2 = |W|^2$ for any input vector V (the interferometric combinations preserve total flux), $U$ is a complex unitary matrix.

The nulling device representation adopted in this study is universal, as the matrix $U$ can describe any coherent mixing scheme between the beams. Any matrix $U$ can be implemented by finite numbers of beam splitters and phase shifters, but the model is not restricted to specific phase shifts or split ratios between beams, as previous studies have sometimes assumed (for example, the Laurance nulling interferometers described in \cite{2000SPIE.4006..871K} are limited to 0 or $\pi$ phase shifts). Proof that any unitary matrix U can be realized with a finite number of beam splitters and phase shifters is given in section 3.3.3 of \cite{2006ApJS..167...81G}. An example beam combiner design is given in \S\ref{ssec:ptsource}, along with a step-by-step description of the beam combiner construction process using phase plates and beam splitters.

\section{Optimal interferometric combinations for a given aperture geometry}
\label{sec:optc}

The optimal solution for a beam combiner aimed at high contrast imaging is a function of the source observed (especially relative position of the planet to the star), so we must carefully define the performance criteria adopted to define what is an optimal beam combiner. We choose to assume that the position of the planet is not known, and that its detection, in the photon-noise limit, is equivalent to the detection of a constant background on the sky (there is no preferred location for the planet). Such a background is incoherent in the nulling interferometer as all possible phases are averaged, and thus its intensity distribution at the output of the interferometer is not affected by the coherent beam combinations. 
For simplicity, we first assume that $U$ is a square matrix and the interferometer consists of equally sized apertures. The total planet intensity $I_p$ gathered by the interfeometer is then equally distributed among the $M$ outputs. The intensity on each interferometer ouput $m$ is the incoherent sum of the starlight intensity $I(star)_m$ and the planet intensity $I_p/M$. While $I(star)_m$ is a function of the matrix $U$, $I_p/M$ is not (planet modeled as incoherent background). The overall detection signal-to-noise (SNR) in the photon noise limit can be written as the quadratic sum of the SNRs on each output:
\begin{equation}
SNR^2 = \sum_{m=0}^M (SNR_m)^2  \propto  \sum_{m=0}^M \left(\frac{I_p/M}{\sqrt{I(star)_m + I_p/M}}\right)^2.
\end{equation}  
With the notation $x_m = I(star)_m/Ip$, 
\begin{equation}
\label{equ:snr2}
SNR^2 = \frac{I_p}{M} \sum_{m=0}^M \frac{1}{1 + M x_m}.
\end{equation}
We note that the sum of the $x_m$ values over $m=0..M-1$ is the ratio of total starlight over total planet light, and is independent of $U$. The nulling beam combiner design can only affect how this sum is spread among the outputs $m=0..M-1$. Maximizing the signal-to-noise ratio given by equation \ref{equ:snr2} is thus achieved by concentrating as much starlight as possible in as few interferometer outputs as possible, therefore leaving other outputs sufficiently dark for detection of high contrast source(s). Equation \ref{equ:snr2} indeed shows that any modification of the interferometer design that moves starlight from output $i$ (decreases $x_i$) to ouput $j$ (increases $x_j$) increases the SNR if and only if $x_j > x_i$.

We can therefore compute the optimal set of interferometric combinations between the interferometer input beams by iterating it: we first direct as much of the starlight as possible in a single coherent output, therefore minimizing the total amount of starlight in all other outputs. The beam combinations between the remaining outputs are then optimized to maximize residual starlight is a single coherent output, and so on. This iterative approach ensures that, provided that the interferometer output beams are ranked in decreasing order of residual starlight, the amount of starlight in the last $k$ outputs is optimally minimal for any value of $k$. The resulting interferometer design will maximize planet detection signal-to-noise ratio (SNR), as it will provide as many nulled outputs as possible, therefore simultaneously optimizing null depth and throughput. This approach is different from only optimizing null depth, which may lead to a design for which a single output is nulled while all other outputs retain a significant amount of starlight - a solution which would offer a low efficiency for planet detection. While optimizing null depth on one output will yield a good signal-to-noise ratio on that output, a single interferometer output only contains a small fraction of the planet light: with N equally sized apertures and N outputs, the fraction of the planet light in a single output is 1/N when averaged over all possible planet positions. Directing as much light as possible to a small number of outputs ensures that the fraction of planet light used for detection is maximized, as several dark outputs will be generated, if made possible by the array geometry and stellar angular size. Another advantage of this approach is that the optimal nulling device design (matrix U) is largely independent of the target parameters (such as the stellar angular size $r_s$, the planet to star position or the contrast), and is only a function of the array geometry, as discussed in this section. 

\subsection{Optimal beam combiner for an unresolved central source}
\label{ssec:ptsource}

If the central source is unresolved, starlight entering the interferometer is fully described by the vector $V$ in equation \ref{equ:vectvk} with $\alpha=0$ and $\beta=0$. If the first line $U_{m=0}$ of matrix $U$ is chosen to be equal to this vector $V(0,0)$, then all starlight will be directed to a single output of the interferometer, and all other outputs will contain no stellar light. This configuration achieves the optimal separation of starlight and planet light, and offers the highest possible performance. 

\begin{figure}
\includegraphics[scale=0.35]{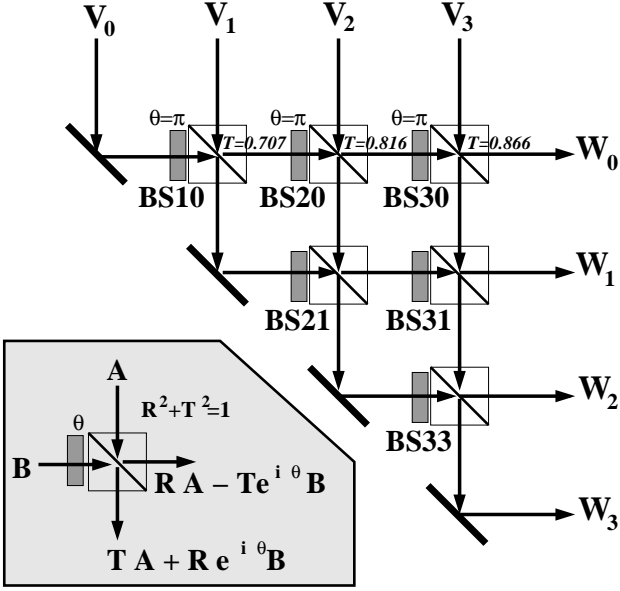} 
\caption{\label{fig:BSc} Possible implementation of a 4-input 4-output nulling beam combiner optimized for an unresolved source. The top row of beam splitters is designed to direct all starlight to ouput $W_0$, leaving the three other outputs dark. The beam splitter unit conventions used in this paper are illustrated in the lower left. Each beam splitter is preceeded by a phase shifting plate, shown in grey, affecting one of its two inputs.}
\end{figure}

For example, for a 4-aperture array with equal subaperture diameters, $V_k = 1/\sqrt{4} = 0.5$ for $k=0,1,2,3$. The optimal beam combiner should have $U_{k,m} = 1/\sqrt{4}$ for output $m=0$. This output can be constructed with 3 beam splitters (BS10, BS20 and BS30 on Fig. \ref{fig:BSc}). Each beam spitter unit $i$ is defined by its amplitude reflectivity $R_i$ (noting that the amplitude transmission $T_i$ is $\sqrt{1-R_i^2}$) and the phase offset $\theta_i$ introduced at its left facing input on Fig. \ref{fig:BSc}. The beam splitter assembly can be designed stepping backwards from the last input to the first input. Satisfying $W_0 = 0.5 V_0 + 0.5 V_1 + 0.5 V_2 + 0.5 V_3$ first imposes $R_{30} = 0.5$, yielding $T_{30} = \sqrt{3}/2$. Mixing the proper amount of input $V_2$ into ouput $W_0$ requires $0.5 = R_{20} T_{30} e^{i(\pi+\theta_0)}$, which yields $R_{20}=1/\sqrt{3}$ and $\theta_{30}=\pi$. Mixing the proper amount of input $V_1$ into output $W_0$ requires $0.5 = R_{10}T_{20}T_{30} e^{i(\theta_{20}+\theta_{30})}$, yielding $\theta_{20}=\pi$ and $R_{10}=1/\sqrt{2}$. Finally, $\theta_{10}=\pi$, and one can verify that $T_{10} T_{20} T_{30} = 0.5$.
The remaining 3 beam splitter units (BS21, BS31 and BS33) can be designed following the same steps to realize any pre-determined matrix $U$, although, in this example, since all starlight is directed to output $W_0$, nulling does not put any requirements on these last 3 beamsplitters.

\subsection{Partially resolved central source}

If the central source is partially resolved, it is no longer described as a single vector V, but as a set of $N_{pt}$ vectors $V_{pt} = V(\alpha_{pt},\beta_{pt})$ with $\alpha_{pt}$ and $\beta_{pt}$ ($pt = 0 .. N_{pt}-1$) uniformly distributed over the stellar area. The measured intensities at the output of the interferometer are obtained by summing incoherently the intensity vectors for each of the points on the stellar surface.
\begin{eqnarray}
\label{equ:Istar}
I_{star} = \frac{1}{N_{pt}} \times \sum_{pt=0}^{N_{pt}-1} I_{pt} \nonumber \\
= \frac{1}{N_{pt}} \times \sum_{pt=0}^{N_{pt}-1} |U V(\alpha_{pt},\beta_{pt})|^2
\end{eqnarray}

The optimal set of interferometric combinations (defined by the matrix $U$) is the one that concentrates the most starlight in the smallest number of outputs. While it is no longer possible to concentrate all starlight in a single output, it is possible to identify the values of the coefficients on the first line of $U$ which maximizes the amount of starlight in the first output ($m=0$) of the interferometer. A small amount of residual starlight is then spread over the other outputs $(m=1..N-1)$ according to the other lines of $U$. The second line of $U$ can be chosen to concentrate as much as this residual starlight on output $m=1$. This process is repeated $N$ times until the full matrix $U$ is built. This iterative process ensures that the stellar flux decreases as rapidly as possible as the interferometer output number increases.

Mathematically, the operations described above form a Principal Components Analysis (PCA), which can be performed by singular value decomposition of the $N_{pt}$-by-$N$ transpose $A$ of the data matrix $A^T$. $A$'s columns are the vectors $V_{pt}$ (each column pt of the matrix $A$ is the vector $V_{pt}=V(\alpha_{pt},\beta_{pt})$):
\begin{equation}
A = U \Sigma B^* 
\end{equation}
where $U$ is a $N$-by-$N$ complex unitary matrix, $\Sigma$ is a $N$-by-$N_{pt}$ diagonal matrix containing the singular values of the $A$ ordered in decreasing amplitude, and $B$ is a $N_{pt}$-by-$N_{pt}$ matrix. The PCA decomposition $Y^T$ of the data matrix $A^T$ is given by:
\begin{equation}
\label{equ:PCAsol}
Y^T = A^T U 	
\end{equation}
The columns of the matrix $U$ are the principal components vectors, ordered in decreasing order of amplitude. The first column of $U$ is the first principal component: it is a unity norm vector which represents the dominant variability (maximum variance) in the data set composed of the vectors $V_{pt}$. The second column of $U$ is the second principal component: it is also a unity vector, orthogonal to the first principal component vector, which represents the dominant variability in the data set after removal of the first principal component contribution in the data. 

Numerical derivation of the optimal beam combinations (matrix $U$) requires the number of points $N_{pt}$ used to model the stellar disk to be larger than the number of apertures $N$ in the interferometer. The solution matrix $U$ in equation \ref{equ:PCAsol} becomes independent of $N_{pt}$ for $N_{pt} >> N$.

\subsection{Interferometer modal response to finite angular stellar diameter}
\label{ssec:modal}

The SVD operation described above identifies and sorts the dominant "modes" $U_m$ $(m=0..N-1)$ present in the set of $N_{pt}$ vectors $V_{pt}$ used to model the star, and builds the interferometric combinations which link each of these modes to a single output of the interferometer. The first mode $U_{m=0}$ is equal to the vector $V(0,0)$ and has a singular value close to 1 (strictly equal to 1 if the star has a radius $r_s = 0$), and subsequent singular values are much smaller unless $r_s$ becomes comparable to the interferometer's diffraction limit.

In a randomly chosen 2-D interferometer with a stellar size small compared to the narrowest fringe spacing, the second and third modes are $2^{nd}$ order modes of stellar size: their intensity (square of amplitude) contribution to a vector $V(\alpha,\beta)$ increases as a linear combination of $\alpha^2$ and $\beta^2$. The next 3 modes are $4^{th}$ order, and the following 4 modes are $6^{th}$ order. This property is a direct consequence of the Taylor expansion of equation \ref{equ:vectvk}:

\begin{eqnarray}
\label{equ:taylorexp2D}
e^{i(\alpha x+\beta y)} = 1 + i \alpha x + i \beta y - \frac{1}{2}(\alpha^2 x^2 + 2 \alpha \beta x y + \beta^2 y^2) \nonumber\\
- \frac{i}{6}(\alpha^3 x^3 - 3 \alpha^2 \beta x^2 y - 3 \alpha \beta^2 x y^2 - \beta^3 y^3) + ... 	
\end{eqnarray}

This expansion, when inserted into equation \ref{equ:Istar}, produces a single $0^{th}$ order mode, two $2^{nd}$ order modes, three $4^{th}$ order modes, followed by modes of $6^{th}$ order or higher. For $r_s$ smaller than the interferometer diffraction limit, these modes correspond to the interferometer outputs when the beam combiner design is optimized: output $m=0$ is the bright $0^{th}$ order mode where most of the starlight is directed, outputs $m=1,2$ are the $2^{nd}$ order modes (light in these outputs of the interferometer increases as the square of stellar size), outputs $m=3,4,5$ are the $4^{th}$ order modes, and so on up to the number of apertures (= number of outputs). The expected correspondence between the modes obtained by Taylor expansion and the modes produces by the SVD is physically due to the very rapid decrease of starlight present in the Taylor expansion modes as a function of mode order. For a small angular radius star, the optimal way to concentrate light in a single interferometric output is to match this output with the $0^{th}$ order mode obtained by the Taylor expansion. The next optimal two modes should then be matched to the $2^{nd}$ order modes in the Taylor expansion, and so on.

\subsection{Important predictions}

The analytical model presented in Section \ref{ssec:modal} can be used to predict key behaviour and limits of nulling interferometers, imposed by the aperture geometry. Some of them are presented and discussed in this section.

\paragraph{ PREDICTION 1: Null order}
{\bf In a random 2-D interferometer of $N$ apertures, the maximum achievable null order $\gamma$ is $\gamma=2$ for $N=2,3$, $\gamma=4$ for $N=4,5,6$, $\gamma=6$ for $N=7,8,9,10$. For any aperture geometry, there exists a coherent mixing of the beams that will reach this null order using a finite number of beam splitters and phase shifters.}

The proof for this prediction is given in Section \ref{ssec:modal}, where it is shown that a SVD decomposition can be used to design the interferometer nulling device reaching the null order stated above. The performance of 2-D nulling interferometers is therefore not a smooth function of number of apertures $N$ when performance is limited by stellar angular size. Increasing the number of apertures from $N=5$ to $N=6$ does not bring a large increase in performance, while going from $N=6$ to $N=7$ does by allowing a $6^{th}$ order null.

\paragraph{PREDICTION 2: Effect of array linearity on null depth}
{\bf For the same number of apertures $N$ equal or greater than 3, 1-D interferometers can reach a higher null order than 2-D interferometers.}

The analysis in Section \ref{ssec:modal} shows that at high contrast, the key to designing a high sensitivity interferometer with a limited number of apertures is to remove the number of relevant low order terms (make the coefficient in front of the term very small) in equation \ref{equ:taylorexp2D} to rapidly gain access to higher order modes without having to increase the number of apertures. This can best be done with a 1-D interferometer, where the Taylor expansion becomes:
\begin{equation}
e^{i(\alpha x)} = 1 + i \alpha x - \frac{\alpha^2 x^2}{2} - \frac{i\alpha^3 x^3}{6} + ... 	
\end{equation}

In any 1-D interferometer, starlight on the interferometer output $m$ (numbered from 0 to $N-1$) therefore increases as the $2m$ power of stellar angular size, allowing a higher order null than a 2-D interferometer with the same number of apertures. Table \ref{tab:minnull} gives the maximum null order achievable as a function of number of apertures for 1-D and 2-D geometries, according to the SVD analysis and Taylor expansion presented in Section \ref{ssec:modal}. For the same number of apertures, considerably higher order nulls can be achieved with 1-D geometry than can be achieved with a 2-D geometry. For example, any 1-D interferometer geometry can provide a $6^{th}$ order null with only 4 apertures, while a randomly chosen 2-D geometry would require 7 apertures. With a small number of apertures, it is thus expected that 1-D or quasi 1-D geometry will be preferred when stellar angular size drives the interferometer's sensitivity.

Previously published interferometer designs illustrate the fundamental advantage of 1-D arrays for null order, and many previously suggested interferometer designs reach the minimum achievable null order shown in table \ref{tab:minnull}. The Angel Cross design \citep{1990ngst.conf...81A} is a 2-D geometry achieving a $\theta^4$ null, while a linear array with the same number of apertures can offer a $\theta^6$ null \citep{1997ApJ...475..373A}. The 2-D 6-aperture Mariotti configuration \citep{2005Icar..178..570M} achieves a $\theta^4$ null. Collapsing the 4-aperture 2-D Angel cross geometry into a 3-aperture 1-D array yields the degenerate Angel cross, with a $\theta^4$ null. The best reported null orders with a small number of apertures are indeed 1-D geometries, and reach the limits shown in table \ref{tab:minnull}: $\theta^6$ null with 4 apertures \citep{1997ApJ...475..373A} and $\theta^8$ null with 5 apertures \citep{1997ASPC..119..285W}.

\cite{2006dies.conf..213R} proposed a method to design 1-D nulling interferometers offering a $\theta^{2L}$ null with $2L$ apertures. The iterative method employed is simple and showed for the first time that arbitrarily high null orders can be achieved given a sufficient number of apertures. As shown by table \ref{tab:minnull}, it is however not optimal, as a $\theta^{2L}$ null can be obtained with any set of $L+1$ apertures.

\begin{deluxetable}{lccl}
\tabletypesize{\scriptsize}
\tablecolumns{3} 
\tablewidth{0pc} 
\tablecaption{\label{tab:minnull} Minimum achievable null order} 
\tablehead{
\colhead{Number of} & \multicolumn{2}{c}{Minimum null order} \\
\colhead{apertures}   & \colhead{1D}   & \colhead{2D} }

\startdata 
\hline
2 & $\theta^2$ & $\theta^2$ \\
3 & $\theta^4$ & $\theta^2$ \\
4 & $\theta^6$ & $\theta^4$ \\
5 & $\theta^8$ & $\theta^4$ \\
6 & $\theta^{10}$ & $\theta^4$ \\
7 & $\theta^{12}$ & $\theta^6$ \\
8 & $\theta^{14}$ & $\theta^6$ \\
9 & $\theta^{16}$ & $\theta^6$ \\
10 & $\theta^{18}$ & $\theta^6$ \\
\enddata 
\end{deluxetable}

\paragraph{PREDICTION 3: Interferometer effective throughput}
{\bf The effective throughput of a nulling interferometer increases with the number of apertures}

The analysis presented in Section \ref{ssec:modal} can also predict the sensitivity of interferometers for high contrast observations. For example, if a given observation requires a $\theta^4$ null, and a 5-aperture 1-D interferometer is used, then at most 3 out of 5 beam outputs can be used toward detection, yielding an effective throughput (averaged over all positions of the planet relative to the interferometer pointing) of 60\%. In this example, the optimal 5-aperture interferometer will have one bright output (where most of the starlight is directed) and one second-order null output, with all remaining outputs being $\theta^4$ or deeper. 

\paragraph{PREDICTION 4: Phase shifts in beam combiner}
{\bf The optimal nulling coherent beam combiner only uses phase shifts equal to 0 and $\pi$}

As discussed in Section \ref{ssec:modal}, in the small stellar size limit, the eigenvectors which represent the contribution of complex amplitudes at the entrance of the interferometer to its outputs match the terms of the Taylor expansion in equation \ref{equ:taylorexp2D}. For example, the term in $\alpha$, which should be used to construct a second-order null output, is the vector $[ixk]$, where $x_k$ is the x-coordinate of aperture number $k$. Equation \ref{equ:taylorexp2D} shows that all terms of the expansion are in the form $(i^n) R$, where $i=\sqrt{-1}$, n is an integer, and $R$ is a vector of real numbers (this is due to the fact that the spatial coordinates x and y are real numbers). Alternatively, each eigenvector can be written as $e^{i \phi} R$, where $\phi$ is 0 or $\pi/2$. The overall phase of the eigenvector has no effect on the beam combiner design, as the intensity output of the interferometer for the set of complex amplitudes $e^{i\phi} R$ at the entrance of the interferometer is independent of $\phi$. When mathematically performing the SVD to identify eigenvectors, pairs of eigenvectors are produced with identical amplitudes and a $\pi/2$ phase offset between the two vectors. Since the two vectors in each pair are physically identical for the interferometer, only one of the two is kept toward designing the nulling interferometer, and its phase can arbitrarily be chosen such that the eigenvector is real. A positive real coefficient in the vector means that the light is not phase-shifted between the input and output, while negative coefficients indicate a $\pi$ phase shift. This prediction is numerically verified by the analysis in this paper, as the optimal nuller design (matrix $U$) is always real. Phases are therefore omitted in the matrix $U$ notations in the rest of the paper, and a minus sign is used for a $\pi$ phase shift.

\section{Interferometer designs}
\label{sec:designs}

\subsection{Linear Arrays}

\subsubsection{Angel \& Woolf 4-aperture design}

\begin{figure*}
\includegraphics[scale=0.3]{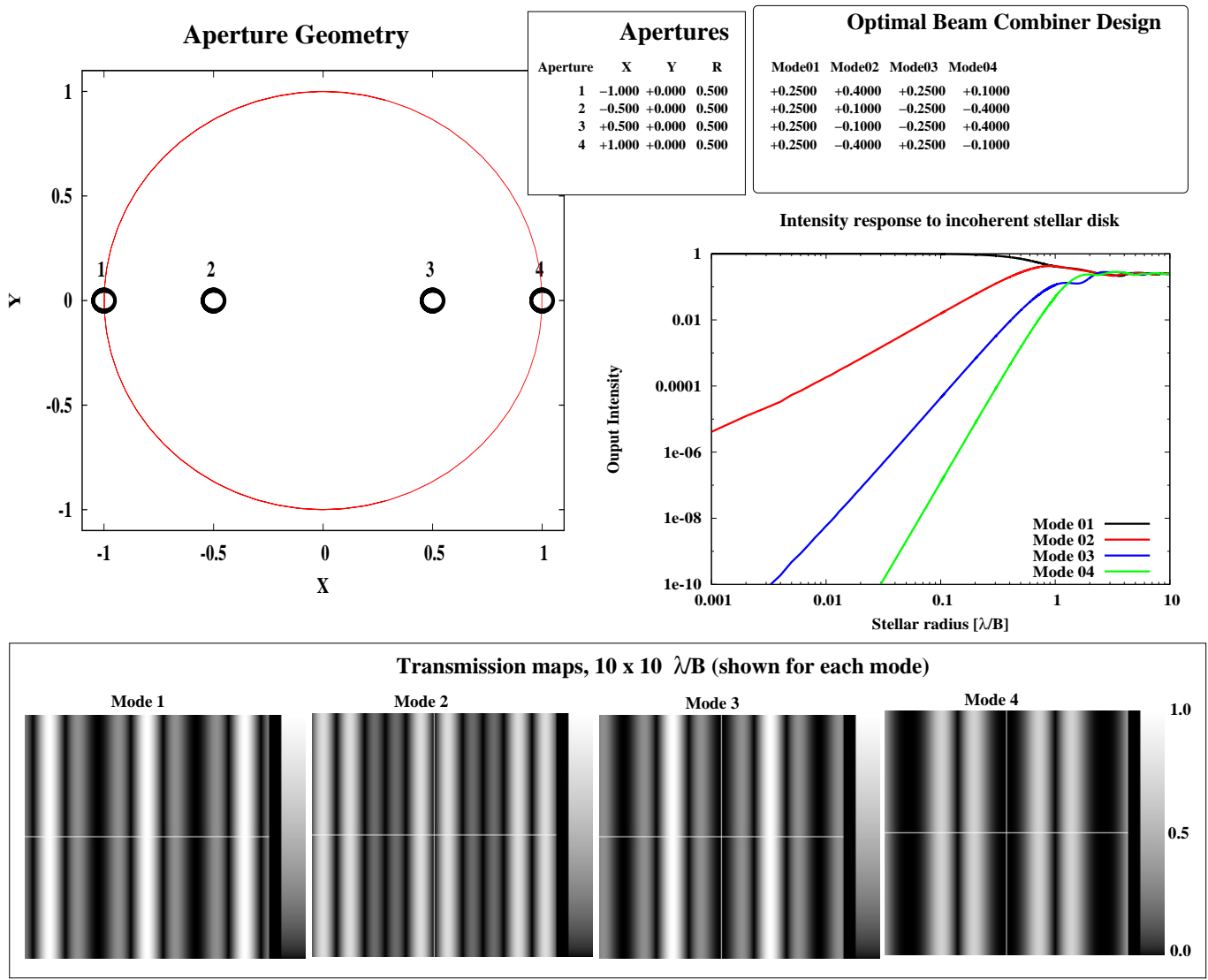} 
\caption{\label{fig:AW97} Beam combiner solution for the linear 4-aperture geometry (top left) proposed in \cite{1997ApJ...475..373A}. In the optimal nulling beam combiner design table (top right), the element $(i,j)$ indicates the fraction of the intensity collected by aperture $(i)$ that is directed to interferometer output $(j)$ (also denoted mode $j$), and a negative value indicates a $\pi$ phase shift. Center right: Distribution of light intensity among the four outputs for the observation of an incoherent stellar disk as a function of stellar angular radius (x-axis). Bottom: Intensity transmission map for each of the four interferometer outputs. The interferometer baseline B is defined in this work as the diameter of the smallest circle containing the interferometer - this circle is shown along with the interferometer geometry (upper left).}
\end{figure*}

\cite{1997ApJ...475..373A} proposed a linear 4-aperture nulling interferometer design offering a $\theta^6$ null. The four equal sized telescopes are arranged in two Bracewell interferometers of baselines $B/2$ and $B$ (in this section, the longest baseline is denoted $B$). The dark outputs of each pair are combined to produce a single $\theta^6$ null. Figure \ref{fig:AW97} shows the result of the SVD approach to designing the optimal beam combiner for this array geometry, assuming a 0.001 $\lambda/B$ stellar radius (where $B$ is the longest baseline) for the central star to ensure that the solution is optimal in the small stellar size limit. We note that this angular radius choice is smaller than typical values for a 20 m to 200 m baseline interferometer observing nearby stars at visible to thermal IR wavelengths, but is adopted here to explore the optimal interferometer design in the small angular size limit. For stellar angular sizes that are much smaller that $\lambda/B$, the interferometer's optimal design is driven to reject as many consecutive terms of the Taylor expansion given in equation \ref{equ:taylorexp2D} as possible, yielding a solution independent of the stellar radius assumed during the numerical optimization.

The first output of the interferometer (Mode 1 in Figure \ref{fig:AW97}) contains most of the starlight, while modes 2, 3 and 4 show respectively $\theta^2$, $\theta^4$, and $\theta^6$ intensity dependence with stellar angular size. The ability to produce a $6^{th}$ order null with this 4-aperture geometry, first described by \cite{1997ApJ...475..373A}, is thus confirmed by the SVD analysis. The beam combinations that produce this null, as given by the SVD (last column of the table at the upper right of Figure \ref{fig:AW97}) also match the combination identified by Angel and Woolf: the dark output of the inner 2-aperture Bracewell interferometer (apertures 2 and 3, coefficients -0.4, 0.4) is combined with the dark output of the outer Bracewell interferometer (apertures 1 and 4, coefficients 0.1, -0.1). As Angel and Woolf described, the two Bracewell interferometers have opposite signs, and the inner interferometer is given 4 times (2 times in amplitude) the weight of the the outer interferometer. The solution given by the SVD analysis offers superior sensitivity to the beam combiner design proposed in \cite{1997ApJ...475..373A}, as it also offers a $\theta^2$ null (output 2) and a $\theta^4$ null (output 3), while the beam combining configuration proposed in \cite{1997ApJ...475..373A} offered two bright outputs (O1 and O2 in fig 1 of their paper) and a single $\theta^2$ output (O3 in fig 1 of their paper). This could be achieved by recombining together outputs O1, O2 and O3 of the beam combiner shown in fig 1 of \cite{1997ApJ...475..373A} according to the values given in the table at the upper right of Figure \ref{fig:AW97}.

\begin{figure*}
\includegraphics[scale=0.3]{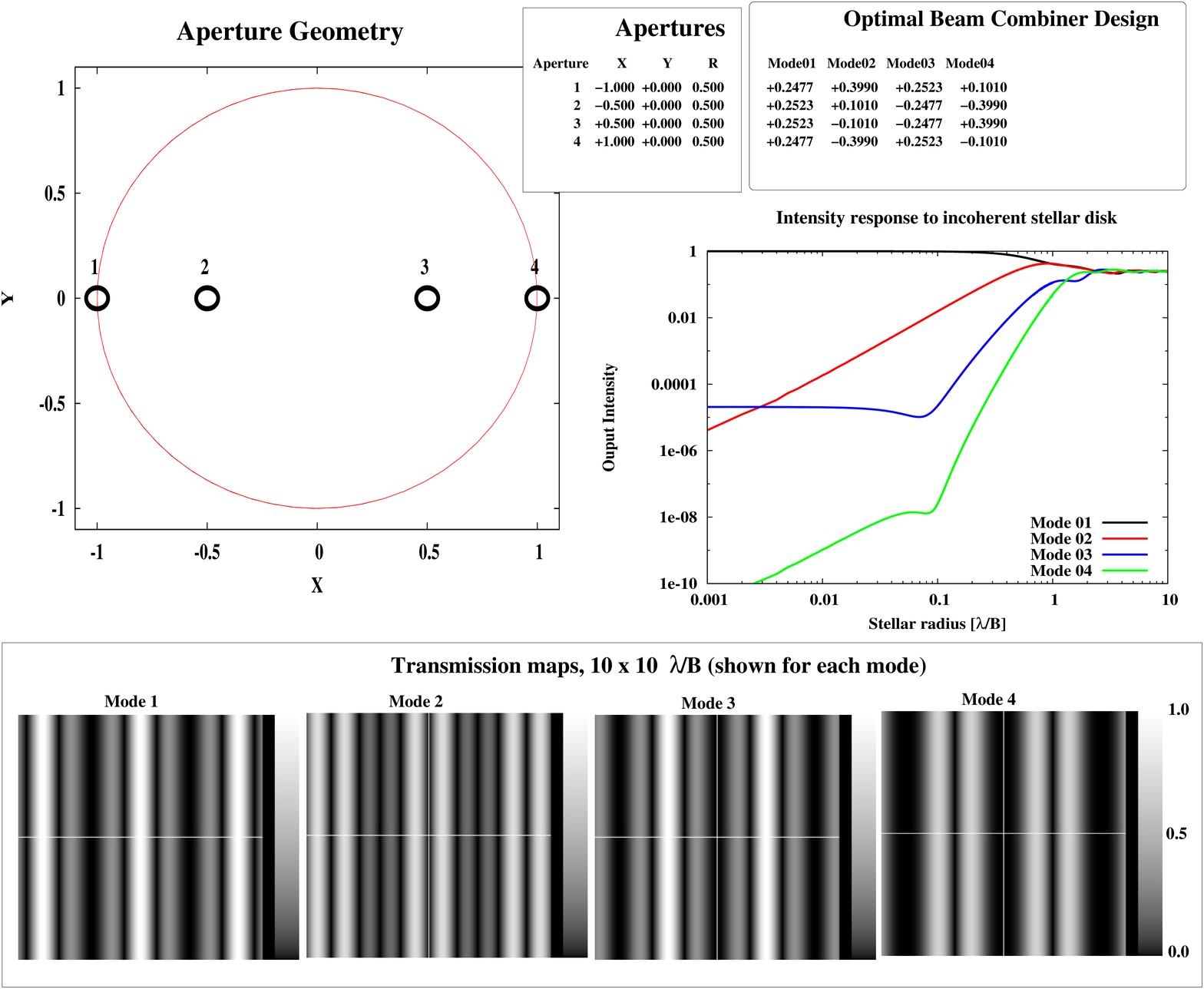} 
\caption{\label{fig:AW97_2} Optimal beam combiner solution for the linear 4-aperture geometry proposed in \cite{1997ApJ...475..373A}, computed here for a 0.1 $\lambda/B$ stellar radius. The solution differs from the small stellar size limit (Figure \ref{fig:AW97}) in the same way as predicted in \cite{1997ApJ...475..373A}: more weight should be given to outer apertures in the deepest null output.  }
\end{figure*}

\cite{1997ApJ...475..373A} suggested slightly increasing the amplitude of the outer Bracewell pair relative to the inner pair to achieve a wider null, suggesting a 0.504 relative ratio in amplitude (as opposed to 0.5 in the design described above). The effect of doing so on the null is shown in fig 2 of their paper, where three local minima in output intensity are shown within the null, located at approximately -0.1 $\lambda/B$, 0 and +0.1 $\lambda/B$. To explore this possibility, the SVD analysis is repeated on the same interferometer geometry, but with an 0.1 $\lambda/B$ stellar radius. Results, shown in Figure \ref{fig:AW97_2}, not only confirm that increasing the relative weight of the outer apertures increases null depth, but demonstrate that doing so is the optimal solution to the nulling beam combiner design. The optimal value for the relative weight in amplitude should be $\sqrt{0.10097/0.39903} = 0.503$ for a 0.1 $\lambda/B$ radius stellar disk, close to the 0.504 value proposed by \cite{1997ApJ...475..373A}, and the transmission map has 3 local minima at -0.8 $\lambda/B$, 0 and +0.8 $\lambda/B$, as shown by the local minimum of the intensity output curve vs. stellar diameter for mode 4 in Figure \ref{fig:AW97_2}.

This last example demonstrates the power of the SVD approach to optimal nulling beam combiner design when stellar size becomes too large for the simple Taylor expansion approximation to remain valid. In this case, the optimal solution starts to differ from the small stellar size limit solution, and the beam combiner is optimally chosen to cancel the incoherent stellar disk in ways that would be difficult to anticipate without the SVD analysis.

\subsubsection{Optimal beam combiner design for arbitrary linear geometries}

A key result of the SVD-based construction of optimal beam combiners for nulling interferometers is the ability to produce deep nulls regardless of aperture geometry. The technique predicts for example that $\theta^6$ and $\theta^8$ nulls can be constructed out of respectively any 4-aperture linear array and any 5-aperture linear array.

\begin{figure*}
\includegraphics[scale=0.3]{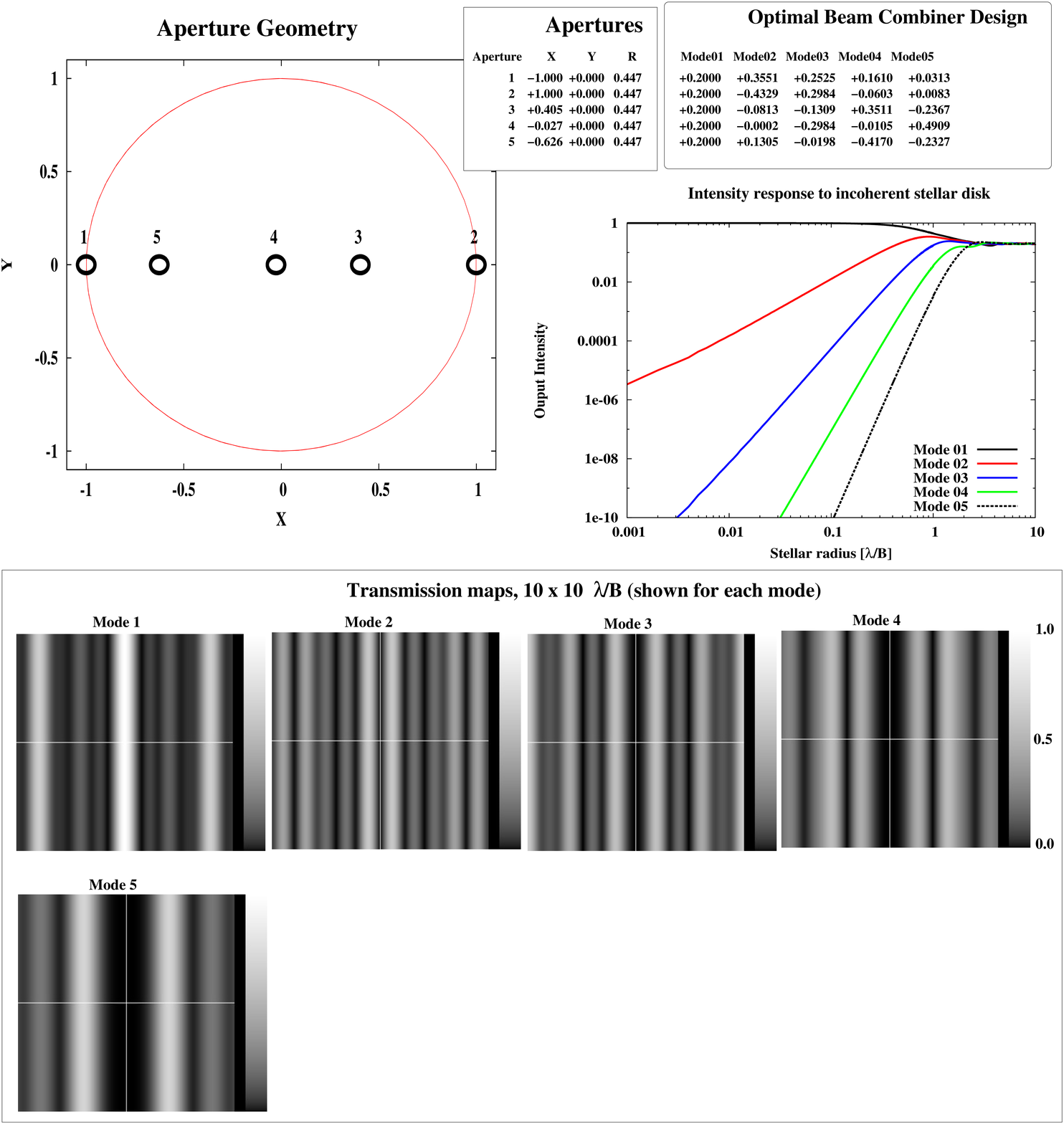} 
\caption{\label{fig:5AR} 
Optimal beam combiner solution for a randomly chosen linear 5-aperture geometry, computed here for a 0.001 $\lambda/B$ stellar radius. As predicted in Section \ref{sec:optc}, the solution produces an output with a deep $\theta^8$ null, along with 3 other nulled outputs offering $\theta^6$, $\theta^4$ and $\theta^2$ deep nulls.}
\end{figure*}

Figure \ref{fig:5AR} shows the result of the SVD-based technique for a randomly chosen 5-aperture linear array geometry with equal aperture sizes. The $\theta^8$ deep null produced is especially resilient to stellar angular size and pointing errors: its transmission for a 0.1 $\lambda/B$ radius disk is below $10^{-10}$. For this randomly chosen geometry, the beam mixing ratios producing the deep null are non-trivial values, with no recognizable integer ratio between the contributions of entrance apertures. The optimal solution also offers other nulled outputs, with nulled orders of respectively 2, 4 and 6, in agreement with the small stellar size limit analysis presented in Section \ref{sec:optc}. 

\begin{figure*}
\includegraphics[scale=0.3]{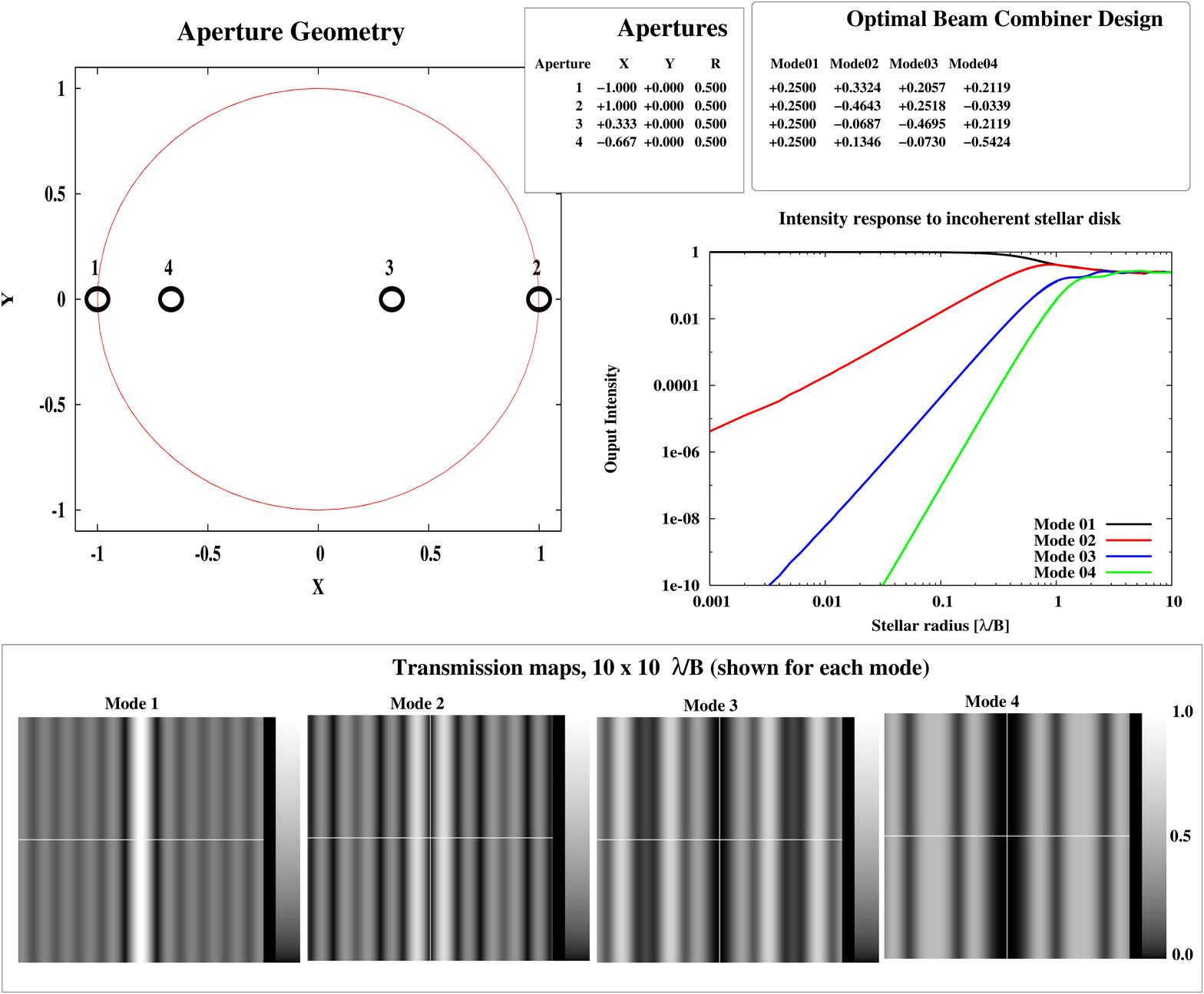} 
\caption{\label{fig:4Auv} 
Optimal beam combiner solution for a linear 4-aperture array geometry optimized for $(u,v)$ plane coverage, computed here for a 0.001 $\lambda/B$ stellar radius. As predicted in Section \ref{sec:optc}, the solution produces an output with a deep $\theta^6$ null, along with 2 other nulled outputs offering $\theta^4$ and $\theta^2$ deep nulls.}
\end{figure*}

The ability to produce deep nulls regardless of aperture geometry allows high performance nulling interferometry to be carried on an array geometry optimized for $(u,v)$ plane coverage. Figure \ref{fig:4Auv} shows the array geometry and beam combiner design for a 4-aperture array optimized for single-dimension $(u,v)$ plane coverage. The baselines covered by this array are all the $bB/6$, with $b=1..6$, allowing excellent $(u,v)$ plane coverage if the array is rotated around the line of sight. The SVD-based technique does produce a $\theta^6$ deep null, along with two other nulls of order 2 and 4 respectively. 

\subsection{2-D arrays}

A key prediction of this study is that any 6-aperture interferometer can achieve a $\theta^4$ deep null. The Taylor expansion derivation proposed in Section \ref{sec:optc} also states that the optimal beam combiner for the interferometer will produce one bright output, two $\theta^2$ outputs and three $\theta^4$ outputs, so the resulting interferometer will have high efficiency provided that a $\theta^4$ deep null is sufficient. Averaged over all possible positions for a planet, it is expected that 50\% of the planet light can be used toward detection or characterization, since half of the interferometer outputs are $\theta^4$ deep nulls.

\begin{figure*}
\includegraphics[scale=0.3]{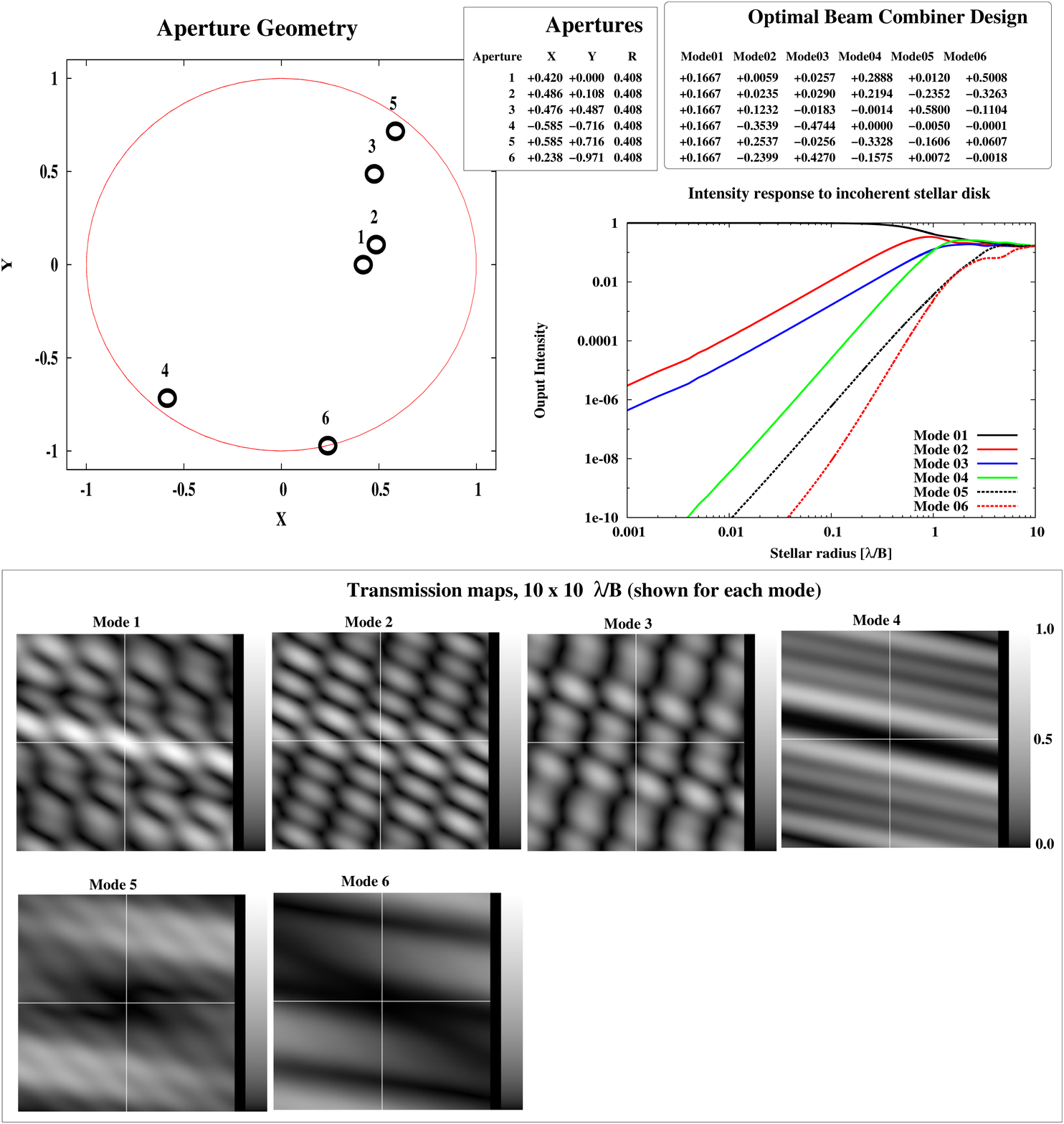} 
\caption{\label{fig:6Auv} Optimal beam combiner solution for a 2-D 6-aperture array geometry optimized for $(u,v)$ plane coverage, computed here for a 0.001 $\lambda/B$ stellar radius. As predicted in Section \ref{sec:optc}, the solution produces three outputs with a deep $\theta^4$ null, along with 2 other nulled outputs offering $\theta^2$ deep nulls.
}
\end{figure*}

Figure \ref{fig:6Auv} illustrates the predictions formulated for the 6-aperture 2-D interferometer. The aperture geometry was first chosen independently of nulling considerations. In this example, the geometry is optimized for $(u,v)$ plane coverage assuming that the interferometer is rotated around the line of sight during the observation. Details of this optimization and the resulting array geometry can be found in \cite{2001PASP..113...98G}. The SVD technique is then used to design the optimal nulling beam combiner, and the results are shown in Figure \ref{fig:6Auv}. As predicted, the beam combiner produces three $\theta^4$ deep nulls, two $\theta^2$ deep nulls, and a bright output which contains almost all of the starlight. Some of the nulled outputs do not offer high transmission within a few $\lambda/B$, so the aperture geometry may not be optimal for imaging and characterization of exoplanets which are likely to be very close to the optical axis. This is due to the aperture geometry, which contains a relatively large number of short baselines to optimize $(u,v)$ plane coverage.

\begin{figure*}
\includegraphics[scale=0.3]{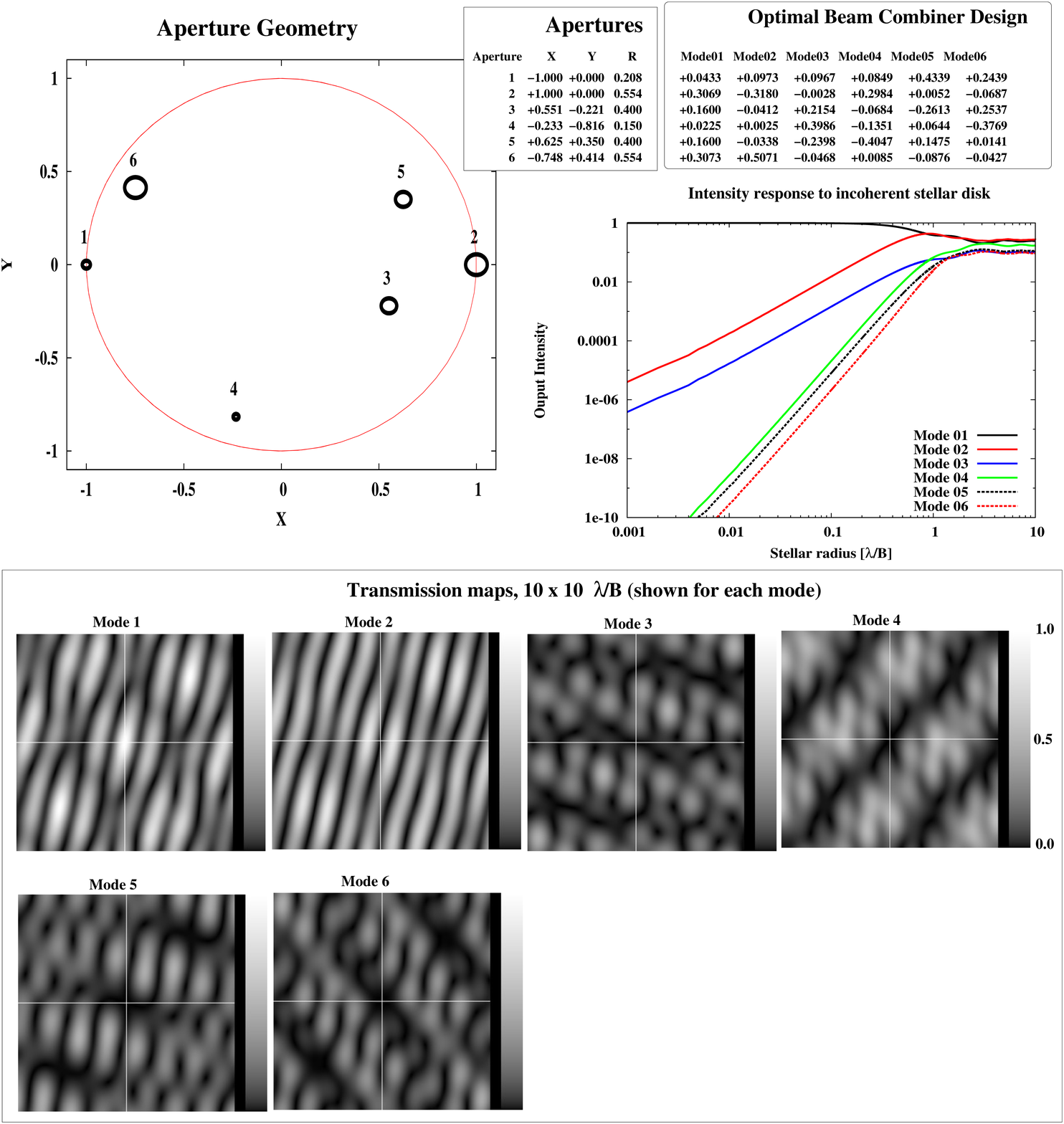} 
\caption{\label{fig:6Ar} 
Optimal beam combiner solution for a 6-aperture array geometry for which the aperture positions and sizes were randomly chosen. The solution was computed for a 0.001 $\lambda/B$ stellar radius. As predicted in Section \ref{sec:optc}, the solution produces three outputs with a deep $\theta^4$ null, along with 2 other nulled outputs offering $\theta^2$ deep nulls. }
\end{figure*}

As illustrated in Figure \ref{fig:6Ar}, the SVD-based technique described in this paper is also applicable to interferometers consisting of uneven aperture sizes, and the findings of this paper also apply to these arrays.

\section{Conclusion}

The analysis performed in this paper shows that the null order in a nulling interferometer is primarily a function of the number of apertures, and that nulls of sufficient depth to detect and characterize exoplanets can be achieved with relatively small number of apertures (4 to 10), regardless of geometry. The superiority of 1-D arrays for achieving deep nulls with a small number of apertures, a trend that is strongly supported by previously published nulling interferometer designs, has been demonstrated and quantified. The SVD-based approach introduced in this paper allows optimal design of beam combiners for nulling interferometry, and is highly flexible, as it can be applied to any geometry, and can also optimally take into account stellar angular size.

While the SVD-based technique has been used to derive the minimal null depth achievable as a function of number of apertures, no strict limit has been placed on the maximum null depth achievable as a function of number of apertures. Specific array geometries may allow deeper nulls than the lower limits shown in table 1, which would enable cost-effective nulling interferometer consisting of very few apertures to be implemented for imaging and spectral characterization of exoplanets.

While the beam combiner designs in this paper are entirely driven by null depth for a finite stellar angular size and planet light throughput, additional considerations must be taken into account in the design of a nulling interferometer, such as sensitivity to background light (especially important at long wavelength, for which zodiacal background exceeds planet light contribution), imaging performance, and resilience to cophasing errors. The impact of exozodiacal light was not considered in this study, but may also drive the optimal array geometry \citep{2010A&A...509A...9D} and beam combining scheme. In some cases, the beam combiner design which is optimal in the null depth sense may not be desirable for a real mission. In particular, a nuller with purely real amplitudes (0 or $\pi$ phase shifts) will produce a centrally symmetric transmission map, making exo-zodi and one-sided point-like signatures difficult to distinguish in the general case, e.g. for a system seen at non zero inclination. The analysis presented in this paper should therefore be extended to include constraints and requirements other than null depth in order to design a beam combiner.

\acknowledgments
The author thanks Frantz Martinache and Vincent Coude-du-Foresto for discussions which helped shape this work. 

\bibliography{ms}

\end{document}